\documentclass[epj]{svjour}
\usepackage{amssymb}
\usepackage{graphics}
\usepackage{psfrag}
\usepackage{epsfig}
\def\bea{\begin{eqnarray}}
\def\eea{\end{eqnarray}}
\def\be{\begin{equation}}
\def\ee{\end{equation}}

\def\La{\Lambda}
\def\Si{\Sigma}

\begin{document}
 \title{Lambda-nuclear interactions and hyperon puzzle in neutron stars}
\author{{J. Haidenbauer\inst{1,2}}
 \and {U.-G. Mei{\ss}ner\inst{2,1}}
 \and {N. Kaiser\inst{3}}
 \and {W. Weise\inst{3}}
}
 \institute{{Institute for Advanced Simulation, Institut f{\"u}r Kernphysik and
 J\"ulich Center for Hadron Physics, Forschungszentrum J{\"u}lich, D-52425 J{\"u}lich, Germany}
\and 
{Helmholtz Institut f\"ur Strahlen- und Kernphysik and Bethe Center
  for Theoretical Physics, Universit\"at Bonn, D-53115 Bonn, Germany}
\and 
{Physik Department, Technische Universit\"at M\"unchen, D-85747 Garching, Germany}
}

\abstract{
Brueckner theory is used to investigate the in-medium properties of a
$\Lambda$-hyperon in nuclear and neutron matter, based on hyperon-nucleon interactions derived within
SU(3) chiral effective field theory (EFT). It is shown that the resulting $\Lambda$
single-particle potential $U_\Lambda(p_\Lambda =0,\rho)$ becomes strongly repulsive for densities $\rho$ of 
two-to-three times that of normal nuclear matter. 
Adding a density-dependent effective $\Lambda N$-interaction constructed
from chiral $\Lambda NN$ three-body forces increases the repulsion 
further. Consequences of these findings for neutron stars are discussed.
It is argued that for hyperon-nuclear interactions with properties such as those deduced from the SU(3) EFT potentials, 
the onset for hyperon formation in the core of neutron stars 
could be shifted to much higher
density which, in turn, could pave the way for resolving the so-called hyperon puzzle.
}

\maketitle 

\section{Introduction}
The interactions of hyperons ($\Lambda$, $\Sigma$, $\Xi$) with nucleons (\(N\)) 
have been in the focus of studies for a variety of reasons~\cite{Gal:2016}. A 
prominent one that has attracted wide attention recently is connected with the 
role that hyperons might play for the mass and size of neutron 
stars~\cite{Chatterjee:2015,Oertel:2016,Tolos:2016}. The observation of 
$2M_\odot$ neutron stars~\cite{Demorest2010,Antoniadis2013} sets strongly restrictive
constraints for the possible appearance of hyperons in neutron star matter and,
accordingly, for the in-medium properties of hyperons and the hyperon-nucleon (\(YN\)) 
interaction itself \cite{Hell2014,Lonardoni2015,Vidana2015,Bombaci:2016}. In order to stabilize such massive 
objects against gravitational collapse a sufficiently stiff equation-of-state (EoS) 
is required which does not leave much room for the presence of hyperons in the 
dense neutron star cores.
A naive introduction of $\Lambda$-hyperons as an additional baryonic degree of
freedom softens the EoS such that it fails to support $2M_\odot$ neutron 
stars \cite{Djapo2010}. 
This is what is commonly referred to as {\it the hyperon puzzle}.
 
Repulsive $\Lambda$-nuclear forces at high baryon densities are considered as a possible 
way to resolve this puzzle. Conventional microscopic calculations using only two-body interactions as input, such as  
Brueckner-Hartree-Fock \cite{Vidana2015,Bombaci:2016} or auxiliary-field diffusion 
Monte Carlo \cite{Lonardoni2015} approaches, do not yield sufficient 
repulsion at high densities. 
Therefore, in the past purely phenomenological mechanisms have been invoked 
that allow to generate the necessary repulsion, e.g.\ through ad-hoc vector 
meson exchange \cite{Weissenborn2012,Weissenborn2012a}, multi-Pomeron exchange 
\cite{Yamamoto2014} or a suitably adjusted repulsive $\Lambda NN$ three-body interaction 
\cite{Lonardoni2015,Vidana:2010}.

Recently, we have presented $YN$ two-body potentials based on SU(3) chiral effective field 
theory 
(EFT) \cite{Polinder:2006,Haidenbauer:2013}. The latest one, derived up to next-to-leading
order (NLO) in the chiral expansion \cite{Haidenbauer:2013}, provides an excellent 
description of the $\Lambda N$ and $\Sigma N$ scattering data. Moreover, it produces a 
satisfactory value for the hypertriton binding energy and decent results for the 
four-body hypernuclei ${}^4_\Lambda {\rm H}$ and ${}^4_\Lambda {\rm He}$ \cite{Nogga:2013}. 
While applications to medium and heavy hypernuclei still remain to be done, 
there have been investigations of the in-medium properties of hyperons for this $YN$ 
two-body interaction, based on a conventional $G$-matrix calculation. 
These studies revealed that (at saturation density) the 
strength of the single-particle potential of a $\Lambda$ is of the order $U_\Lambda = -(25\dots 30)$ 
MeV \cite{Haidenbauer:2014,Petschauer:2016M}, in agreement with empirical information 
deduced from the binding energies of heavy $\Lambda$-hypernuclei \cite{Millener88,Yamamoto88}. 
Moreover, the $\Sigma$-nuclear potential was found to be repulsive, in accordance with the available 
phenomenology \cite{Gal:2016}.  

The in-medium studies in Refs.\,\cite{Haidenbauer:2014,Petschauer:2016M} have focussed on densities 
around normal nuclear matter density, $\rho_0\simeq 0.16$~fm$^{-3}$. 
However, the results for $U_\La$ shown in \cite{Haidenbauer:2014,Petschauer:2016M} 
already indicated that the chiral EFT potentials at NLO could have quite different properties
at higher densities than those from previous one-boson exchange $YN$-inter\-actions. In 
particular, an onset of repulsive effects is seen around $\rho_0$ 
corresponding to a Fermi momentum of $k_F\simeq 1.34$ fm$^{-1}$ in symmetric nuclear matter. 

In the present paper we reconsider this issue and extend the $G$-matrix calculations 
to higher densities. As will be reported below, it turns out that the repulsive effects 
increase dramatically with rising densities. Already at densities $\rho \sim (2-3)\rho_0$, where hyperons might 
appear in the inner core of neutron stars according to the aforementioned studies 
\cite{Lonardoni2015,Vidana2015,Bombaci:2016,Djapo2010}, $U_\Lambda$ is basically repulsive. 

A further new aspect to be studied here in the context of SU(3) chiral EFT are effects
from three-baryon forces. Recently, a density-dependent effective
baryon-baryon interaction has been deduced from (irreducible) chiral SU(3)-based three-baryon forces 
\cite{Petschauer:2017}, in a scheme consistent with the chiral $YN$ two-body 
interaction \cite{Petschauer:2016}. The derivation was done in close analogy to the work of 
Ref.~\cite{Holt:2009} where density-dependent corrections to the \(N\!N\)-interaction were 
calculated from the leading-order chiral three-nucleon forces.
The effective baryon-baryon interaction is obtained from the three-baryon interaction by 
closing two nucleon lines into a loop, diagrammatically representing the sum over occupied states 
within the Fermi sea. An exploratory evaluation \cite{Petschauer:2017} showed that the density-dependent 
$\La N$-inter\-action deduced from the irreducible $\La NN$ three-body forces is repulsive. In the present 
paper we examine this effect more quantitatively by including it in a $G$-matrix calculation. 

\section{Formalism}
The present study is performed using conventional {Brueckner} theory. We summarize 
below only the essential elements. A more detailed description can be found in 
Refs. \cite{Haidenbauer:2014,Reu94}, see also Ref.\,\cite{Vid00}.
We consider a $\Lambda$ (or $\Sigma$) hyperon with momentum ${\vec p}_Y$ in
nuclear or neutron matter at density $\rho$. In order to determine the
in-medium properties of these hyperons we employ the Brueckner reaction-matrix
formalism and calculate the $YN$ reaction matrix $G_{YN}$, defined by the 
Bethe-Goldstone equation
\begin{eqnarray}
\nonumber
&&\langle YN | G_{YN}(\zeta) | YN \rangle = \langle YN | V | YN \rangle  
+\sum_{Y'N} \ \langle YN | V | Y'N \rangle \\
&&\quad\times 
\langle Y'N | \frac{Q}{\zeta - H_0}|Y'N \rangle \ \langle Y'N | G_{YN}(\zeta) | YN \rangle , 
\label{Eq:G1}
\end{eqnarray}
with $Y$, $Y'$ = $\La$, $\Si$.
Here, $Q$ denotes the Pauli projection operator  which excludes intermediate
$YN$-states with the nucleon inside the Fermi sea.
The starting energy $\zeta$ for an initial $YN$-state with momenta ${\vec p}_Y$ and 
${\vec p}_N$ is given by
\begin{equation}
\zeta = E_Y (p_Y) + E_N (p_N),
\end{equation}
where the single-particle energy $E_\alpha (p_\alpha)$ ($\alpha = \Lambda, \Sigma, N$)
includes not only the (nonrelativistic) kinetic energy and the baryon mass but
in addition the single-particle potential $U_\alpha (p_\alpha, \rho)$:
\begin{equation}
E_\alpha (p_\alpha) = M_\alpha + \frac{\vec p^{\,2}_\alpha}{2M_\alpha} + U_\alpha (p_\alpha,
\rho)\, .
\label{Eq:G2}
\end{equation}
The conventional 'gap-choice' for the intermediate-state spectrum is made. Using the 'continuous choice' instead, it was shown that the resulting $\Lambda$-nuclear potential depth differs by less than $2\%$ from the 'gap-choice' calculation \cite{Petschauer:2016M}.

The $\La$ single-particle potential $U_\La(p_\La,
\rho)$ is given by the following integral 
and sum over diagonal $\La N$ $G$-matrix elements:
\begin{equation}
U_\La(p_\La,
\rho) = \int\limits_{|\vec p_N|< k_F}  {d^3p_N\over (2\pi)^3}\, \rm{Tr}
\langle {\vec p}_\La ,{\vec p}_N | G_{\La N}(\zeta) | {\vec p}_\La, {\vec p}_N \rangle 
\,,
\label{Eq:G3}
\end{equation}
where  $\rm{Tr}$ denotes the trace in spin- and isospin-space.
Note that $\rho = 2k^3_F/3\pi^2$ for symmetric nuclear matter and
$\rho = k^3_F/3\pi^2$ for neutron matter.
Eqs.\,(\ref{Eq:G1}) and (\ref{Eq:G3}) are solved self-consistently in the standard way, 
with  $U_\La(p_\La,
\rho)$ appearing also in the starting energy $\zeta$.
As in Ref.~\cite{Haidenbauer:2014} the nucleon single-particle potential 
$U_N(p_N,\rho)$ is taken from a calculation of pure nuclear or neutron matter employing a 
phenomenological $N\!N$-potential. Specifically, we resort to results for the Argonne $v_{18}$
potential published in Ref.\,\cite{Isaule:2016} which are available up to rather 
high nuclear densities. 
As pointed out in Ref.\,\cite{Reu94}, calculations of the $\Lambda$ and $\Sigma$
hyperon potentials in nuclear matter using the gap-choice are not too sensitive to the 
details of $U_N(p_N,\rho)$. Indeed, the difference for $U_\La(p_\La=0,\rho)$ using 
$U_N(p_N,\rho)$ from Ref.\,\cite{Isaule:2016} or the parameterization utilized in 
Ref.\,\cite{Haidenbauer:2014} amounts to less than 1~MeV at nuclear matter saturation 
density $\rho_0$. 

In the present study we employ the $YN$ two-body potentials derived in 
Ref.\,\cite{Haidenbauer:2013} within SU(3) chiral EFT. Specifically, we use the NLO 
interactions corresponding to the cutoffs $\Lambda$\,=\,450\,MeV and 500\,MeV. Both lead to 
values of $U_\La(0,\rho_0)\approx -30$~MeV in line with empirical information from the 
binding energies of heavy $\Lambda$-hypernuclei. 

Furthermore, we perform calculations in which the additional density-dependent effective 
$\La N$-interaction derived from the leading chiral $\La NN$ three-baryon force 
\cite{Petschauer:2016} is taken into account. Details on its derivation and explicit 
expressions can be found in Ref.\,\cite{Petschauer:2017}. Here we just mention that
there are contributions to this force from two-pion exchange, one-pion exchange 
and a contact term. Two-pion exchange gives rise to a spin-independent (central) 
interaction, and to symmetric and antisymmetric spin-orbit interaction terms, 
while the other two components lead only to central forces. SU(3) flavor symmetry and 
decuplet saturation have been used to estimate the involved coupling constants. 
Specifically, the $\La NN$ three-baryon force is saturated 
via the excitation of the spin-3/2 $\Sigma^*(1385)$ resonance. There are only two parameters
in the resulting density-dependent $\La N$-interaction. One of these is the coupling constant of
octet and decuplet baryons with the pseudoscalar meson-octet, denoted by $C$ in 
Ref.\,\cite{Petschauer:2017}. For this constant we use the large-$N_c$ value $C=3g_A/4$, 
where $g_A=1.27$ is the nucleon axial-vector coupling constant. The other parameter, $H'$ 
in Ref.\,\cite{Petschauer:2017}, is a combination of coupling constants of the four-baryon 
contact terms. General dimensional scaling arguments have been invoked which led to
the estimated range $H'\approx \pm 1/f_\pi^2$, with $f_\pi\approx 92$\,MeV  the pion-decay
constant. In the present study we choose the negative value for $H'$. In this case there is 
a partial cancellation between the one-pion exchange three-baryon force and the contact 
term \cite{Petschauer:2017}. This ensures that, for densities around $\rho_0$, the effects
of the density-dependent $\La N$-interaction are still relatively small so that our results for 
$U_\La(0,\rho)$ at nuclear matter saturation density do not change much and remain consistent with constraints from hypernuclear physics.

For the off-shell extension of the density-dependent $\La N$-interaction we follow
the suggestion of Ref.\,\cite{Holt:2009}. This means that we make the substitution
$p^2 \rightarrow \frac{1}{2}(p'^2+p^2)$, where $p$ and $p'$ are the initial and final
center-of-mass momenta of the baryons. In addition, the high-momentum components of the 
interaction are cut off by a regulator function of the form $f_R= \exp\left[-\left(
p'^4+p^4\right)/\Lambda^4\right]$, when inserted into the $G$-matrix equation. It is the 
same regulator as used for the (free-space) $YN$ two-body interaction in the 
coupled-channel Lippmann-Schwinger equation, see Ref.\,\cite{Haidenbauer:2013}. 

\begin{figure*}[t]
\begin{center}
\includegraphics[height=95mm,angle=-00]{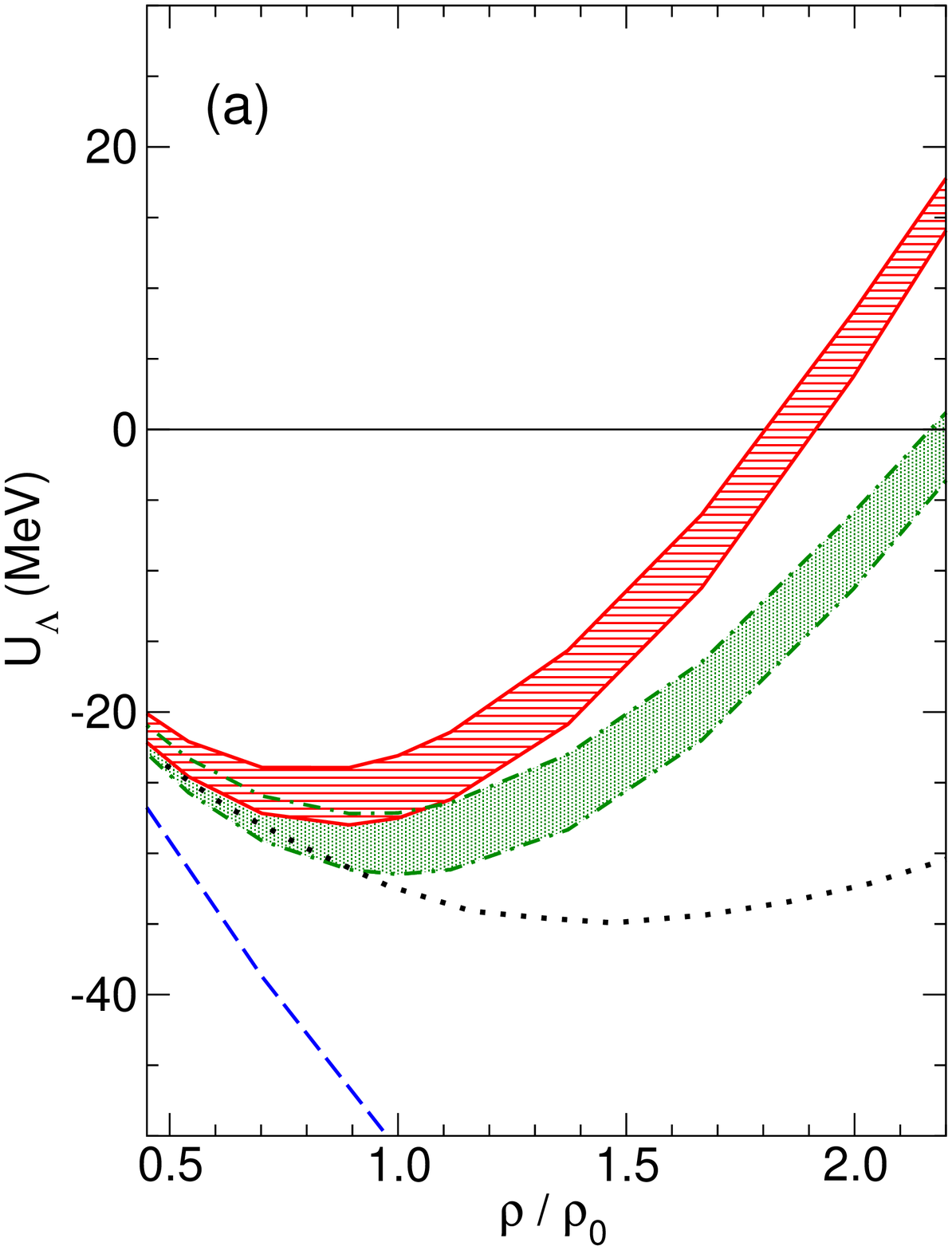}\hspace{0.5cm}
\includegraphics[height=95mm,angle=-00]{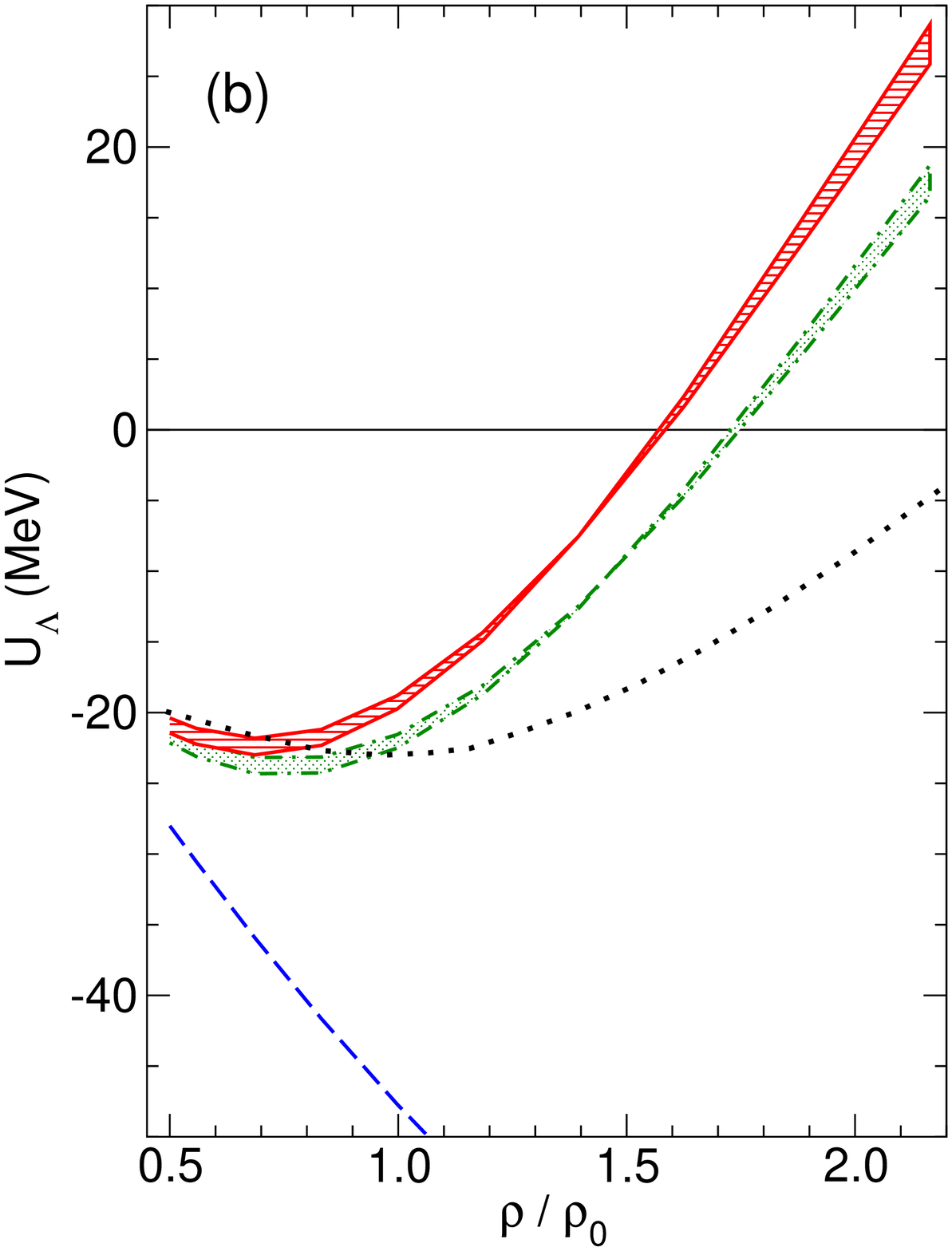}
\vspace{-3mm}
\caption{The $\Lambda$ single-particle potential $U_\Lambda (p_\Lambda = 0,\rho)$ 
as a function of $\rho/\rho_0$ in symmetric nuclear matter (a) and in neutron matter (b). 
The dash-dotted curves show the chiral EFT results at NLO for the cutoffs
$\Lambda =$ 450~MeV (lower curve) and 500~MeV (upper curve), respectively.
The solid lines include the density-dependent $\La N$-interaction derived from the $\Lambda NN$ three-body force \cite{Petschauer:2017}.
The dashed curve is the result of the J{\"u}lich '04 meson-exchange model \cite{Hai05},
the dotted curve that of the Nijmegen NSC97f potential \cite{Rij99}, taken
from Ref.~\cite{Yamamoto:2000}. }
\label{fig:G}
\end{center}
\vspace{-5mm}
\end{figure*}

%
\section{Results and discussion}
Results for the density dependence of the $\Lambda$ single-particle potential are presented
in Fig.\,\ref{fig:G} for symmetric nuclear matter (a) and for neutron matter (b). 
Apart fom predictions from our chiral EFT interactions \cite{Haidenbauer:2013} (dash-dotted lines), 
those for meson-exchange $YN$ models constructed by the J\"ulich \cite{Hai05} 
(dashed line) and Nijmegen \cite{Rij99} (dotted line) groups are also shown.  
As already emphasized in Ref.\,\cite{Haidenbauer:2014}, at low densities the chiral EFT 
potentials exhibit a relatively weak density dependence as compared to that of the 
J\"ulich '04 potential. One observes an onset of repulsive effects around the 
saturation density of nuclear matter, i.e. $\rho = \rho_0$, see dashed-dotted lines in Fig.\,\ref{fig:G}. 
Now, with the calculation extended to higher densities, it becomes clear that these effects 
increase dramatically. Already around $\rho \approx  2\rho_0$, 
$U_\Lambda (0,\rho)$ turns over to net repulsion. 
The NSC97f model \cite{Rij99} exhibits likewise a trend toward repulsion with increasing density. However, the turning
point is at much higher density (dotted line). Other $YN$-interaction models for which 
pertinent results can be found in the literature, like the Nijmegen ESC04 interaction (cf. 
Fig.\,12 in Ref.\,\cite{Rij06}) or an interaction derived within the constituent quark-model 
(fss2) \cite{Fu07} exhibit a trend similar to the one of the J\"ulich '04 potential, i.e. 
a more or less monotonously increasing attraction with rising density. 

Results for $U_\Lambda (0,\rho)$ based on a $G$-matrix calculation that includes the 
density-dependent effective $\La N$-interaction derived from the leading $\La NN$ 
three-baryon forces are shown by solid lines in Fig.\,\ref{fig:G}. One can see that for low 
densities, $\rho/\rho_0\approx 0.5$, the effects of the three-baryon forces are 
essentially negligible. But they become noticeable already around
$\rho=\rho_0$ and, of course, significant at higher density where the repulsion 
strongly increases. 

\begin{figure*}[htb!]
\begin{center}
\includegraphics[height=75mm,angle=-00]{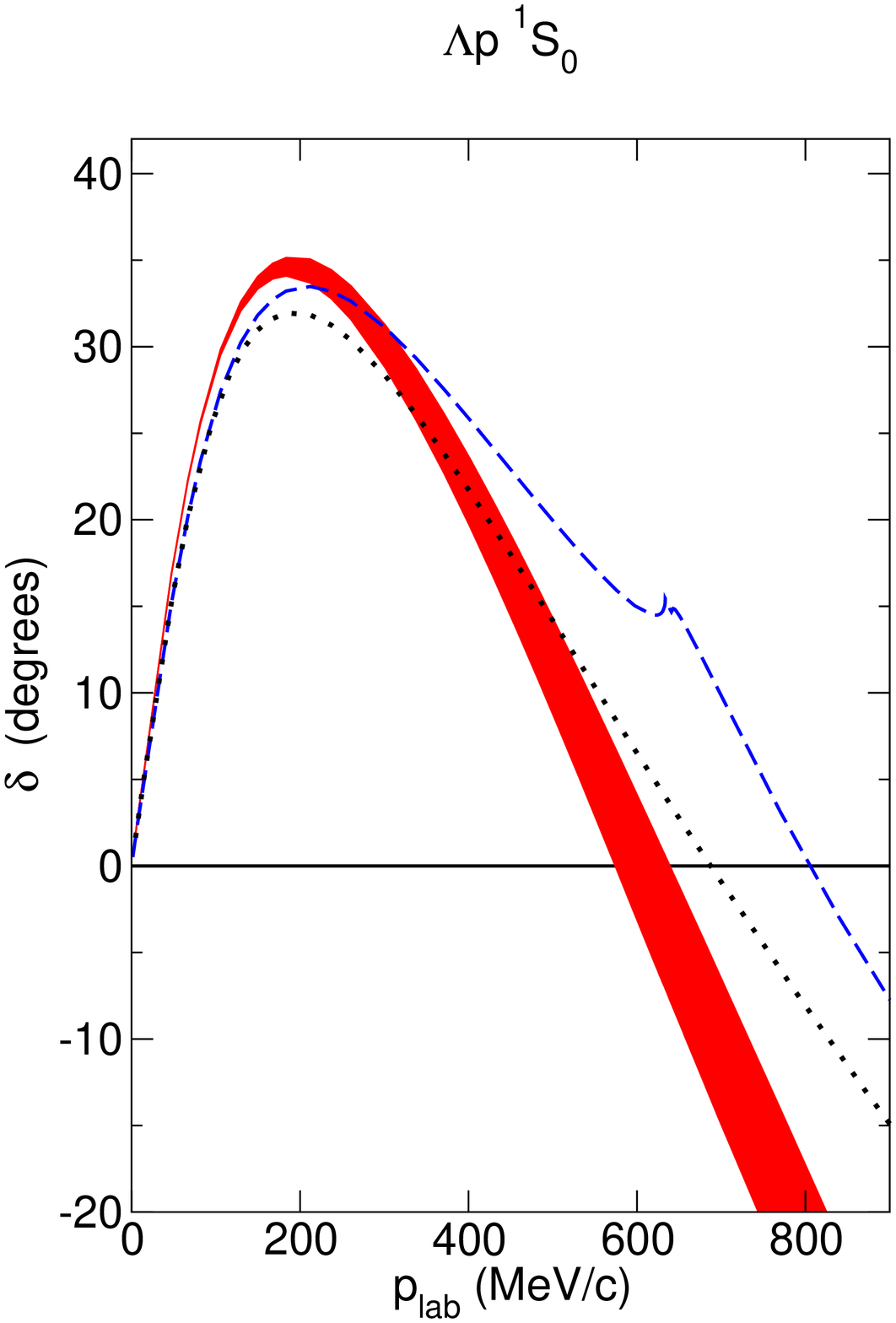}\includegraphics[height=75mm,angle=-00]{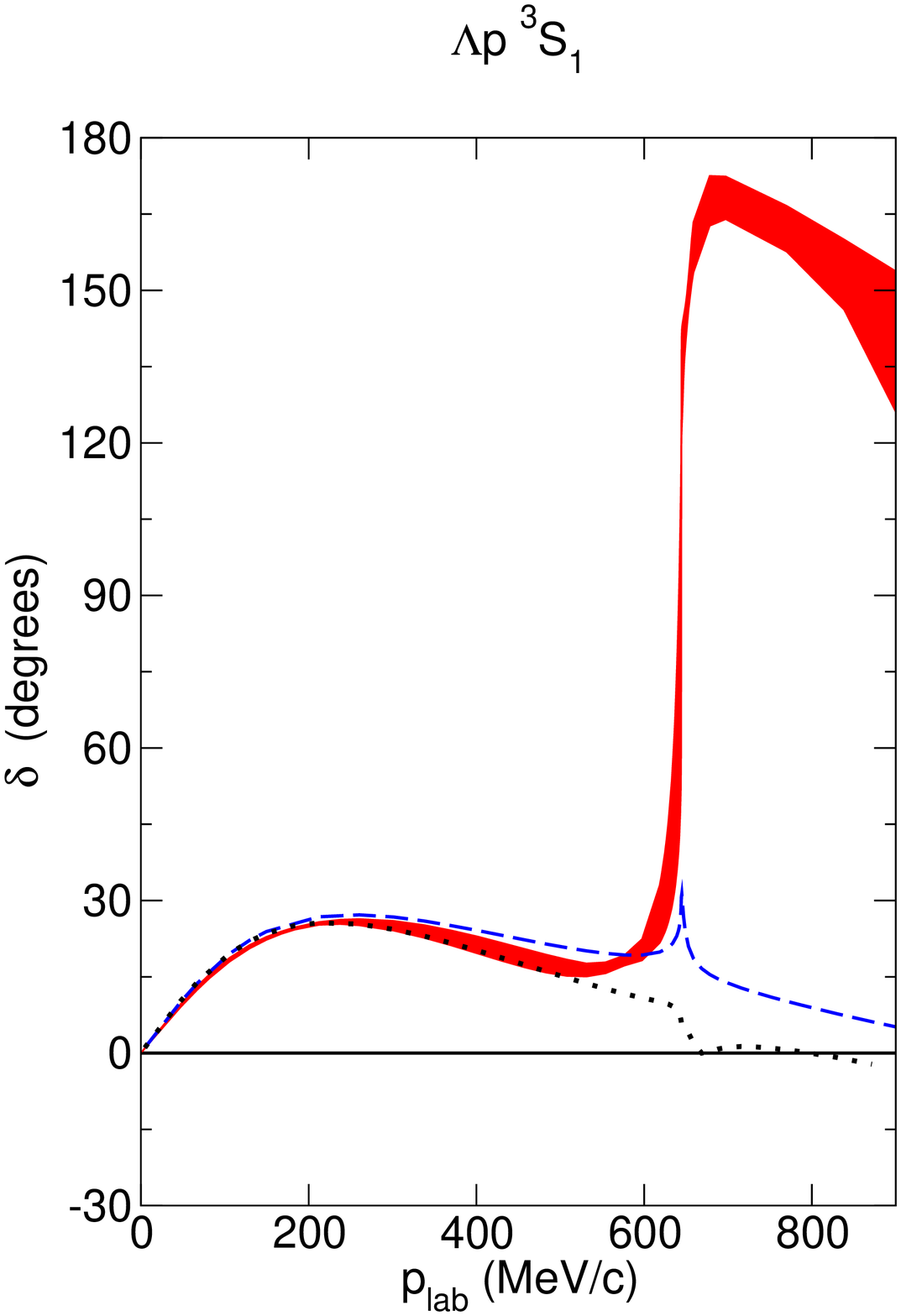}
\includegraphics[height=75mm,angle=-00]{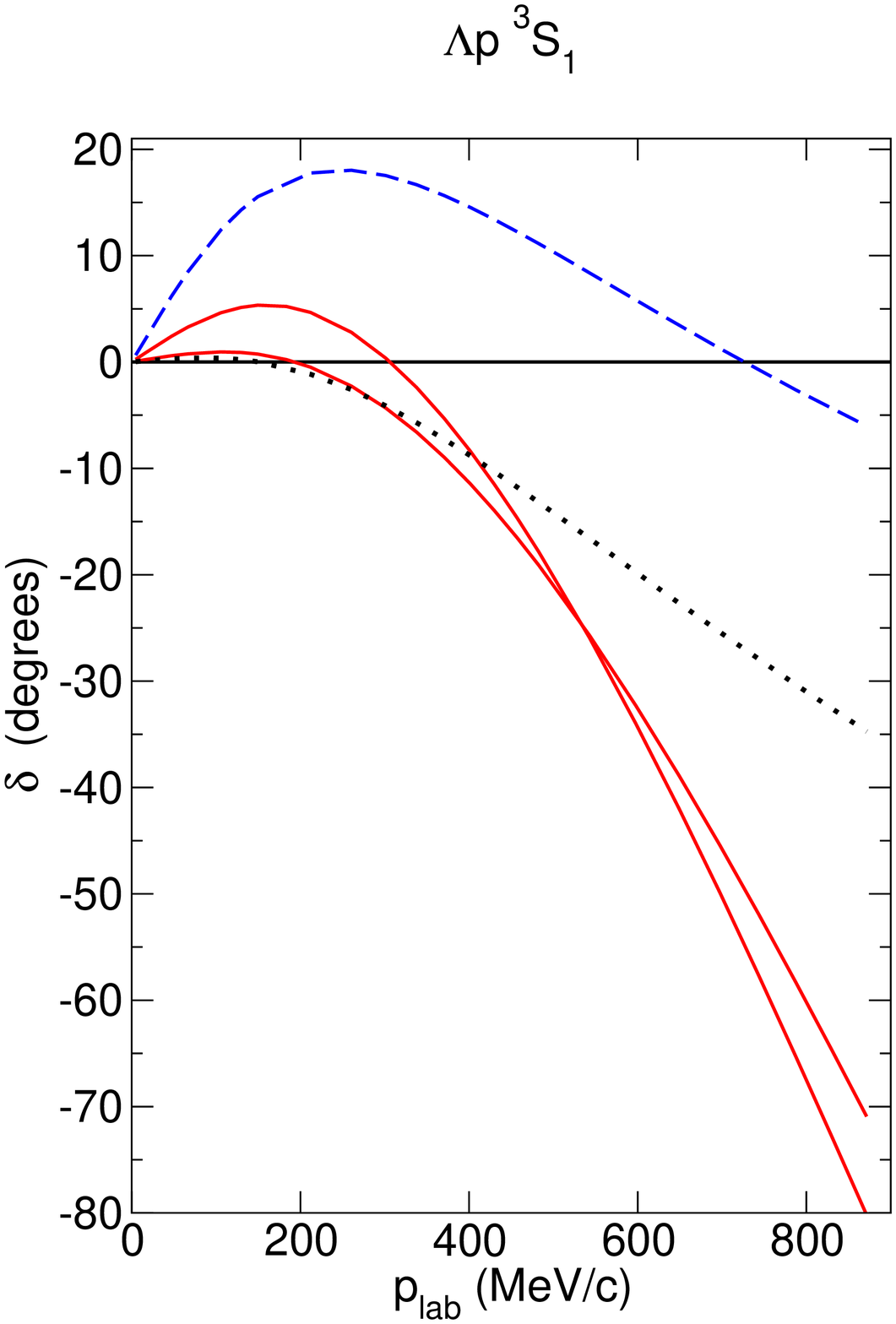}
\vspace{-3mm}
\caption{$\La N$ phase-shifts in the $^1S_0$ and $^3S_1$ partial waves. Full results
including the coupling to $\Si N$, taken from Ref.~\cite{Haidenbauer:2013}, are shown 
in the left and middle panel, respectively.
The red (dark) band is the result for the chiral EFT interaction at NLO. 
The dashed curve corresponds to the J{\"u}lich '04 meson-exchange interaction 
\cite{Hai05}, the dotted curve to that of the Nijmegen NSC97f interaction \cite{Rij99}.
In the right panel results for the $^3S_1$ phase-shift are displayed (for cutoffs 
$\Lambda$\,=\,450\,MeV and 500\,MeV) where the coupling to the $\Si N$-channel is switched off.
}
\label{fig:P}
\end{center}
\vspace{-3mm}
\end{figure*}

How can we understand these results in terms of the properties of the underlying 
$YN$-interactions?
For that we take a look at the $^1S_0$ and $^3S_1$ $\La N$ partial-waves which provide 
the bulk contribution to the single-particle potential $U_\Lambda(p_\La,\rho)$ 
\cite{Haidenbauer:2014,Petschauer:2016M}. 
In the case of the $^1S_0$ partial-wave, see Fig.\,\ref{fig:P} (left), 
the phase-shift computed with the NLO chiral EFT interaction crosses zero at lower momenta 
compared to the Nijmegen NSC97f potential (dotted line), and at much 
lower momentum than the J\"ulich '04 potential (dashed line). This suggests
that the EFT interactions are more repulsive at short distances. 
An inspection of the pertinent contributions to $U_\Lambda(p_\La,\rho)$ reveals,
however, that their density dependence is similar for all potentials considered. 
Thus, these differences at high momenta do not influence the matter properties in a 
qualitative way. 

The situation is different for the coupled $^3S_1$-$^3D_1$ partial wave. Its properties 
are strongly influenced by the coupling between the $\La N$ and $\Si N$ channels 
driven by the one-pion exchange tensor force. This is evident from the $^3S_1$ 
phase-shift~\cite{Haidenbauer:2013}, reproduced here in Fig.\,\ref{fig:P} (middle),
where one observes either a cusp or a rounded step~\cite{Newton,Machner}  
at the opening of the $\Si N$-channel. 
Corresponding results for the $^3D_1$ phase-shift can be found in Refs.~\cite{Polinder:2006,Haidenbauer:2013}.
It is important to realize that the actual strength of the channel coupling differs for the 
different $YN$ interactions as visible in the characteristic differences of the $^3S_1$ 
phase-shift around the $\Si N$ threshold. 
One can see that more directly by simply switching off the $\La N \leftrightarrow \Si N$ 
coupling. Corresponding results are shown in Fig.\,\ref{fig:P} (right). 
Obviously, in the J\"ulich~'04 model the $\La N$-interaction itself is already fairly 
attractive. The $\La N$--$\Si N$ coupling is only moderate as suggested by the fact that 
up to $p_{\rm{lab}} \approx 400$~MeV/c there is not much difference in the phase-shifts
between the coupled and uncoupled calculation. 
Contrary to that, in case of the chiral EFT interactions (but also for the 
NSC97f interaction) the $\La N$-interaction itself is predominantly repulsive 
and, accordingly, the overall attraction reflected in the full $^3S_1$ result
is provided solely by a rather strong $\La N \leftrightarrow \Si N$ coupling. 
Indeed,
in both scenarios the (final) $\La N$ $^3S_1$ phase-shifts (Fig.\,\ref{fig:P}, middle) 
are comparable at momenta $p_{\rm{lab}} \lesssim 600$~MeV/c and of the proper magnitude 
as required to describe the $\La N$-scattering data. 

Clearly, the $\La N$ and $\Si N$ scattering data themselves 
do not allow to discriminate between the scenarios discussed above. 
The situation is quite different, however, when such interactions are employed in 
calculations of hypernuclei~\cite{Gibson:1994,Hiyama:2001,Nogga:2002}
and/or hyperon properties in nuclear matter~\cite{Kohno:2013}.
This is well-known for many years and has been discussed, e.g., in the context of light
hypernuclei in Refs.\,\cite{Gibson:1994,Hiyama:2001,Nogga:2002}. Specifically, in the work
by Gibson et al.\,\cite{Gibson:1994} the effect of $\La N$--$\Si N$ coupling (also called 
$\La$--$\Si$ conversion) has been reviewed and it has been argued that the $\La$--$\Si$ 
conversion in the nuclear medium is suppressed as compared to that in free space.
For a related discussion in the context of neutron matter see Ref.\,\cite{Kohno:2013}.
Accordingly, with regard to our $G$-matrix calculation one expects that at
higher densities the $\La N$--$\Si N$ coupling gets increasingly suppressed.
As a consequence, the in-medium properties are to a greater extent determined by 
the (diagonal) $\La N$-interaction alone. 
If this interaction is weakly attractive or even repulsive as for the NLO chiral EFT
interaction~\cite{Haidenbauer:2013}, $U_\La(0,\rho)$ will become repulsive at 
higher densities. This is precisely what we observe for the EFT interactions
where the contribution of the $^3S_1$ partial-wave to $U_\La(0,\rho)$ changes sign 
as the density increases.
On the other hand, if the $\La N$--$\Si N$ coupling is fairly weak (as for the 
J\"ulich '04~interaction), or when using simple effective $\La N$-potentials which ignore 
the coupling to the $\Si N$-channel altogether, one ends up with a persistently attractive 
in-medium $\La N$-potential. Typically, such interactions generate too strongly attractive 
results for $U_\La(0,\rho)$ in $G$-matrix calculations. Likewise they lead to overbinding 
in few- and many-body calculations of hypernuclei \cite{Nogga:2013,Wirth:2016}.
This deficiency can then be cured only by introducing an ad-hoc 
strongly repulsive phenomenological $\La NN$ three-baryon force 
\cite{Wirth:2016,Lonardoni:2013}. 

Contributions from higher partial waves, specifically from the $P$-waves, 
play an increasingly important role at higher densities. Most of those are repulsive for the NLO 
chiral EFT interactions \cite{Haidenbauer:2014,Petschauer:2016M}, and also for the
Nijmegen NSC97f potential \cite{Rij99,Yamamoto:2000}. 

At this point a discussion of remaining uncertainties is useful and necessary. One might ask whether the NLO treatment of the $YN$ interactions is 
sufficient, given the fact that the chiral EFT approach to the $NN$ interaction itself has reached a level far beyond that approximation. 
However, the available empirical YN data base is still much inferior in quality and quantity to the one provided by the accurate $NN$
phase shifts and bound-state information. The $\Lambda N$ and $\Sigma N$ scattering data, within their present empirical uncertainties, are well reproduced at NLO. 
Going beyond NLO at this stage would just increase the number of parameters, without improvements in precision. 
In particular, most of the low-energy constants of the subleading meson-baryon vertices that would enter at 
NNLO \cite{Epelbaum:2008} are basically unknown for the SU(3) sector. 
On the other hand, the important two-pion exchange 
mechanisms that govern the $\Lambda$-nuclear interaction at the relevant distance scales are generated already at NLO. 
For further progress it would of course be highly desirable to have better hyperon-nucleon data sets.

Concerning $\Lambda NN$ three-body forces derived from chiral SU(3) EFT, we rely on the analysis and estimates performed in 
Ref.~\cite{Petschauer:2017}. In that work the number of free parameters is significantly reduced by assuming SU(3) decuplet dominance in intermediate states of the three interacting baryons, in a way similar to introducing the $\Delta$ isobar as an explicit degree of freedom in the non-strange NNN sector. Then, for example, the $\Lambda NN$ three-body force involving two-pion exchange does not introduce any new parameters, its couplings being related to the corresponding ones in the $NNN$ sector by SU(3) coefficients. This $\pi\pi$ exchange piece of the $\Lambda NN$ interaction produces a density-dependent effective $\Lambda N$ two-body force that is repulsive in all partial waves, irrespective of quantitative details. A remaining unknown parameter, denoted $H'$ in \cite{Petschauer:2017}, is associated with the four-point (contact) vertex $\Lambda N \leftrightarrow \Sigma^* N$.  The three-body $\Lambda NN$ contact term, when 'resolved' and factorized into $\Lambda NN \rightarrow \Sigma^* NN \rightarrow\Lambda NN$, is proportional to $(H')^2$. Its contribution to the equivalent effective $\Lambda N$ two-body interaction is again repulsive, independent of the (unknown) sign of $H'$, and grows linearly with density. In the one-pion exchange piece of the $\Lambda NN$ three-body force the $H'$ parameter appears linearly. For $H' < 0$ there is partial cancellation between this piece and the  $\Lambda NN$ contact term, but the net repulsion from the sum of all 3-body terms survives under any circumstances. In fact, the effects of $H'$ are active only in $S$-waves, while $P$-waves and higher partial waves experience 3-body repulsion exclusively from the $\pi\pi$ exchange process for which the parameters are fixed when assuming decuplet dominance. 

In practice, dimensional arguments \cite{Friar:1996,Epelbaum:2002}
lead to expect a 'natural' order of magnitude $|H'| \sim 1/f^2$, with $f \sim 0.1$~GeV 
the pseudoscalar meson decay constant. Our choice used in the present work, $H' = - 1/f_\pi^2$, corresponds to a 'minimal' 3-body repulsion scenario that introduces small corrections at normal nuclear densities, compatible with empirical constraints from hypernuclei, but still adds significantly to the growing repulsive $\Lambda$-nuclear forces as the baryon density increases.

\section{Conclusions}
The coupling between the $\La N$- and $\Si N$-channels plays an important role in the
hyperon-nucleon interaction. Its strong influence on the properties of light hypernuclei 
has been thoroughly examined and discussed in the past 
\cite{Nogga:2013,Gibson:1994,Nogga:2002}.
The results of $G$-matrix calculations for nuclear and neutron matter reported in the present work reveal 
that this coupling has also a crucial impact on in-medium properties of $\La$-hyperons. 
This conclusion is based on the hyperon-nucleon interaction derived recently within SU(3)
chiral EFT up to NLO and, in contrast, on the J\"ulich~'04 interaction \cite{Hai05} 
as a representative of conventional one-boson exchange $YN$-models. 
The former interaction is characterized by a relatively weak diagonal $\La N$-interaction and a  
strong $\La N$--$\Si N$ coupling, whereas in the J\"ulich~'04 model the $\La N$-interaction 
itself is fairly attractive and, accordingly, the $\La N$--$\Si N$ coupling is much 
weaker. While both forces yield comparable and satisfactory descriptions of the 
available $\La N$ and $\Si N$ scattering data \cite{Haidenbauer:2013,Hai05},  
their predictions for the single-particle potential $U_\La(0,\rho)$ differ qualitatively. 
Specifically, for the chiral EFT interaction the $\Lambda$-nuclear single particle 
potential $U_\La(0,\rho)$ becomes increasingly repulsive at higher densities, whereas the one 
of the J\"ulich '04 model remains strongly 
attractive throughout and does not satisfy the empirical constraints from (heavy) hypernuclei. 

Let us finally discuss implications for neutron stars. 
It should be clear that it is mandatory to include the $\La N$--$\Si N$ 
coupling in the pertinent calculations. This represents a challenging task since 
standard microscopic calculations without this coupling are already quite complex. However, without the 
$\La N$--$\Si N$ coupling, which has such a strong influence on the in-medium properties 
of hyperons, it will be difficult if not impossible to draw reliable conclusions. 
The majority of $YN$-interactions employed so far in microscopic calculations of neutron stars 
have properties similar to those of the J\"ulich '04 model. 
In such calculations, hyperons appear in the core of neutron stars typically at densities 
around $(2-3)\rho_0$ \cite{Lonardoni2015,Vidana2015,Bombaci:2016}. This leads to a 
strong softening of the equation-of-state and consequently to a maximal 
mass of a neutron star far below the $2 M_\odot$ constraint.
Assume now that nature favors a scenario with a weak diagonal $\La N$-interaction 
and a strong $\La N$--$\Si N$ coupling as predicted by SU(3) chiral EFT. The present study demonstrates that, 
in this case, the $\Lambda$ single-particle potential $U_\La(0,\rho)$ based on chiral EFT two-body interactions is 
already repulsive at densities $\rho \sim (2-3)\rho_0$.  The one of the $\Sigma$-hyperon is likewise repulsive 
\cite{Haidenbauer:2014}. We thus expect that the appearance of hyperons in neutron stars will 
be shifted to much higher densities. In addition there is a moderately repulsive density-dependent 
effective $\La N$-interaction
that arises within the same framework from the leading chiral $YNN$ three-baryon forces. It 
enhances the aforementioned repulsive effect and would make the appearance of $\Lambda$-hyperons in 
neutron star matter energetically unfavorable  and unlikely, even at the central densities that can be reached in $2 M_\odot$ stars with radii $R > 10$ km. In summary, all these aspects taken together 
may well point to a  possible solution of the so-called hyperon puzzle without resorting to exotic mechanisms. 

\vskip 0.4cm
{\bf Acknowledgments:}
This work is supported in part by the DFG and the NSFC through
funds provided to the Sino-German CRC 110 ``Symmetries and
the Emergence of Structure in QCD''. The work of UGM was also 
supported by the Chinese Academy of Sciences (CAS) President's
International Fellowship Initiative (PIFI) (Grant No. 2017VMA025).

%

\end{document}